\documentclass[pra,reprint,amsmath,amssymb,aps,superscriptaddress]{revtex4-1}

\usepackage{graphicx}
\usepackage{dcolumn}
\usepackage{bm}

\usepackage{xargs}
\usepackage[pdftex,divipsnames]{xcolor}
\usepackage[colorinlistoftodos,prependcaption,textsize=tiny]{todonotes}
\newcommandx{\tdunsure}[2][1=]{\todo[linecolor=red,backgroundcolor=red!25,bordercolor=red,#1]{#2}}
\newcommandx{\tdchange}[2][1=]{\todo[linecolor=blue,backgroundcolor=blue!25,bordercolor=blue,#1]{#2}}
\newcommandx{\tdadd}[2][1=]{\todo[linecolor=yellow,backgroundcolor=yellow!25,bordercolor=yellow,#1]{#2}}

\newcommand{\ket}[1]{\vert #1 \rangle}
\usepackage{amsfonts}

\usepackage[hidelinks]{hyperref}

\usepackage{cleveref}
\crefname{figure}{Fig.}{Figs.}
\crefname{section}{}{}
\crefformat{equation}{(#2#1#3)}

\begin{document}

\preprint{APS/123-QED}

\title{Measurement-based generation of shaped single photons and coherent state superpositions in optical cavities}

\author{Ruvindha L. Lecamwasam}
\affiliation{Centre for Quantum Computation and Communication Technology, Department of Quantum Science, Research School of Physics and Engineering, The Australian National University, Canberra ACT 2601 Australia}
\author{Michael R. Hush}
\affiliation{University of New South Wales at the Australian Defence Force Academy, Canberra ACT 2600 Australia}
\author{Matthew R. James}
\affiliation{Centre for Quantum Computation and Communication Technology, Research School of Engineering, The Australian National University, Canberra ACT 2601 Australia}
\author{Andr\'e R. R. Carvalho}
\affiliation{Centre for Quantum Computation and Communication Technology, Department of Quantum Science, Research School of Physics and Engineering, The Australian National University, Canberra ACT 2601 Australia}
\date{\today}

\begin{abstract}

We propose related schemes to generate arbitrarily shaped single photons, i.e. photons with an arbitrary temporal profile, and coherent state superpositions using simple optical elements. The first system consists of two coupled cavities, a memory cavity and a shutter cavity, containing a second order optical nonlinearity and electro-optic modulator (EOM) respectively. Photodetection events of the shutter cavity output herald preparation of a single photon in the memory cavity, which may be stored by immediately changing the optical length of the shutter cavity with the EOM after detection. On-demand readout of the photon, with arbitrary shaping, can be achieved through modulation of the EOM. The second scheme consists of a memory cavity with two outputs which are interfered, phase shifted, and measured. States that closely approximate a coherent state superposition can be produced through postselection for sequences of detection events, with more photon detection events leading to a larger superposition. We furthermore demonstrate that `No-Knowledge Feedback' can be easily implemented in this system and used to preserve the superposition state, as well as provide an extra control mechanism for state generation.

\end{abstract}

\maketitle

\section{Introduction} 
Nonclassical states of light are an essential resource in optical quantum information processing. Single photons are ideally suited for transmission along a quantum network~\cite{Sangouard:2007}, are used for secure communication in quantum cryptography~\cite{artBB84,bkQuantumComputingAndInformation}, and with only linear optical components can be used to implement scalable and robust quantum computing \cite{artQuantumComputingLinearReview,artQuantumComputingLinear,artRSISinglePhotonDetectorsReview,bkQuantumComputingAndInformation}. 
Superpositions of coherent states are another important class of nonclassical states which are not only useful for quantum information processing~\cite{artCatStatesUsefulness,Marek:2010}, but also allow for fundamental tests of quantum mechanics and the mechanism of decoherence~\cite{artObservingDecoherence,artReconstructionNonClassicalCavityFieldStates}.  

There has been much experimental effort into generating these states in a variety of physical systems. In the optical domain, the production of propagating coherent-states superpositions (CSS) requires some type of non-Gaussian operation, which can be achieved by hybrid strategies~\cite{NEERGAARD-NIELSEN:2011,Andersen:2015} that combine techniques from the fields of continuous and discrete variable quantum optics~\cite{Neergaard-Nielsen:2006,Ourjoumtsev:2006,Ourjoumtsev:2007,Takahashi:2008,Chrzanowski:2011,Huang:2015}. Schemes have also been implemented in the microwave domain \cite{artSynthesizingArbitraryQuantumStatesSuperconductingResonator,artHofheinz2008}, with the system in \cite{artSynthesizingArbitraryQuantumStatesSuperconductingResonator} deterministically preparing arbitrary quantum states in a superconducting resonator by carefully controlling its interaction with an auxiliary qubit.

In the case of single photons, sources are categorised as either `heralded' or `on-demand'. The former generate photons probabilistically, but signal this production to the observer. These generally use the process of spontaneous parametric downconversion, which can operate at a wide range of frequencies, including those best suited for long-range communication~\cite{artSPGenerationAndDetection}. The resulting states are well-defined with a high level of purity, at the expense of the efficiency of pair production. In contrast there are `on-demand' sources, including quantum dots, NV centres and trapped ions \cite{artRSISinglePhotonDetectorsReview}, which excite a physical energy level that emits a photon as it relaxes, eliminating the probabilistic aspect. In general, these sources present limitations due to the low efficiency for collecting emitted photons, although recent progress in quantum dots~\cite{Somaschi:2016} allowed for the production of a highly pure and bright single photon source. 

There is also interest in generating shaped shaped single photons sources \cite{artIonSPG,Carvalho2012}, where by the `shape' of a single photon we mean the probability distribution of the time of emission. These can allow for exciting atoms and cavities with unit efficiency \cite{artQuantumStateTransfer, Bader:2013,Sondermann:2015}, minimising errors due to mode mismatching in interference experiments \cite{artOptimalPhotonsForQIP}, better transmission along optical fibers \cite{artCavitySPG, artSolitonsFibers}, and ultimately building a framework for quantum information based on photonic temporal modes~\cite{Brecht:2015}. 
 
Inspired by the recent on-demand single photon source demonstrated in an optical setup~\cite{Yoshikawa:2013}, in this article we outline a scheme to produce shaped single photon pulses and CSS using standard optical components and measurement-based feedback. We begin in~\cref{secScheme} with a brief overview of the theory of open quantum systems, and use this to describe a model of the source from~\cite{Yoshikawa:2013}. In~\ref{secSPG} we propose a method for generation of shaped single photons, and analyse its optimal regimes of operation. In \cref{secAnotherChannel} we modify the detection scheme to produce CSS in the optical cavity, and show how one can increase the storage time by implementing `No-Knowledge' feedback \cite{Szigeti:2014}.  

\begin{figure} [t!]
  \includegraphics[width=0.9 \columnwidth]{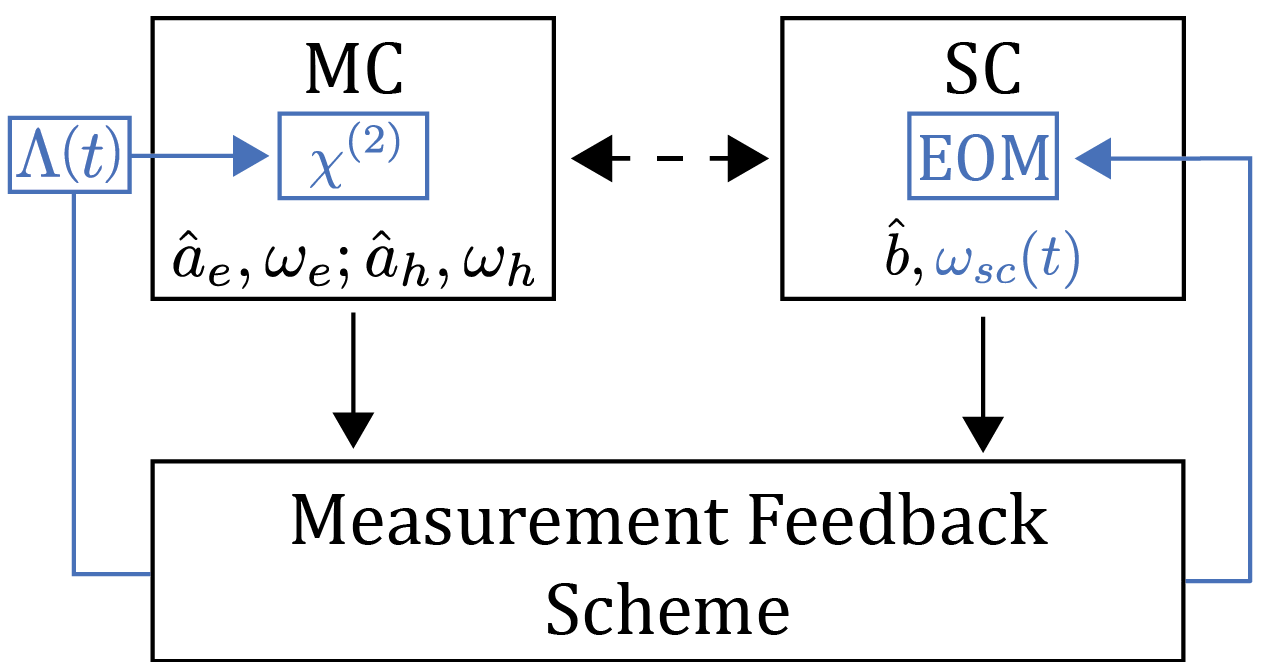}
  \caption{(Colour online) a) The scheme consists of two coupled cavities, MC (Memory Cavity) and SC (Shutter Cavity). The former supports two modes, $\hat{a}_h$ (`heralding') and $\hat{a}_e$ (`emission'), with frequencies $\omega_h$ and $\omega_e$ respectively. Pairs of photons $\hat{a}_h^\dagger\hat{a}_e^\dagger$ are created by pumping of an optical nonlinearity $\chi^{(2)}$ with a laser.  SC supports a single mode $\hat{b}$ of frequency $\omega_{sc}(t)$ which may be tuned by the EOM. The output channels of the two cavities may be monitored in a number of ways, and based on this we vary $\omega_{sc}(t)$ and the pumping $\Lambda(t)$ of $\chi^{(2)}$.}\label{figBlockDiagram}
\end{figure}

\section{Model of the Scheme}
\label{secScheme}

We show in \cref{figBlockDiagram} the general state preparation scheme, consisting of two coupled cavities, Memory Cavity (MC) and Shutter Cavity (SC). MC supports two photon modes, $\hat{a}_h$ and $\hat{a}_e$, called `heralding' and `emission' with frequencies $\omega_h$ and $\omega_e$ respectively, as well as a $\chi^{(2)}$ optical nonlinearity \cite{bkFundamentalsOfPhotonics}, which when pumped with a coherent field $\Lambda(t)$ produces pairs of photons $\hat{a}_h^\dagger\hat{a}_e^\dagger$. The time dependence signifies that the pump field is either switched on or off, depending on the stage of the scheme. We will later use superscripts to distinguish if the nonlinearity is nondegenerate ($\chi^{(2,n)}$), producing modes at different frequencies, or degenerate ($\chi^{(2,d)}$), where the modes have the same frequency but orthogonal polarisations. SC supports a single mode ${\hat b}$ whose frequency $\omega_{sc}(t)$ may be varied using an Electro-Optic Modulator (EOM) \cite{bkFundamentalsOfPhotonics}. The cavity outputs are measured, with the results used to control the EOM and the driving field of the nonlinearity.

The Hamiltonian describing the dynamics of the nonlinear crystal and the interaction between cavity modes is given by
\begin{equation}\label{eqFullHamiltonian}
  \begin{aligned}
    \hat{H}=\;&\hbar\omega_h\hat{a}_h^\dagger\hat{a}_h+\hbar\omega_e\hat{a}_e^\dagger\hat{a}_e+\hbar\omega_{sc}(t){\hat b}^\dagger{\hat b}\\
            &+\hbar\Lambda(t)(\hat{a}_h\hat{a}_e+\hat{a}_h^\dagger\hat{a}_e^\dagger) \\
            &+\hbar g(\hat{a}_h{\hat b}^\dagger+\hat{a}_h^\dagger {\hat b} + \hat{a}_e{\hat b}^\dagger+\hat{a}_e^\dagger {\hat b}),
  \end{aligned}
\end{equation}
where $g$ is the coupling between cavities. To complete our description, we need to include the loss of photons from the cavities, which can be done by considering the master equation~\cite{bkQMC}
\begin{equation}\label{eqMasterEquationSchrodinger}
  \frac{d\hat{\rho}}{dt}=-\frac{i}{\hbar}[\hat{H},\hat{\rho}]+\sum_j\mathcal{D}[\hat{L}_j]\hat{\rho},
\end{equation}
where $\rho$ is the density matrix for the state of the system, $\mathcal{D}$ is the Lindblad superoperator defined as
\begin{equation}\label{eqLindbladD}
  \mathcal{D}[\hat{r}]\hat{\rho}=\hat{r}\hat{\rho}\hat{r}^\dagger-\frac{\hat{r}^\dagger \hat{r}}{2}\hat{\rho}-\hat{\rho}\frac{\hat{r}^\dagger\hat{r}}{2},
\end{equation}
and $\{\hat{L}_j\}$ are operators representing the decoherence processes. Assuming that the photon losses occur for the emission, heralding and shutter cavity modes at rate $\kappa_e$, $\kappa_h$, and $\kappa_{sc}$ respectively, the decoherence operators are
\begin{equation}\label{eqFullLossOperators}
  \begin{aligned}
    \hat{L}_e &= \sqrt{\kappa_{e}}\hat{a}_e, \\
    \hat{L}_h &= \sqrt{\kappa_{h}}\hat{a}_h, \\
    \hat{L}_{sc} &= \sqrt{\kappa_{sc}}\hat{b}. \\
  \end{aligned}
\end{equation}

We can combine and measure the decoherence channels in different ways, and in the following sections we will show how this can be used to generate specific states in the emission mode of MC. 

\subsection{Modelling the detection}

As already mentioned, our state preparation scheme is based on continuous monitoring of the output fields of the cavities. Since Eq.~(\ref{eqMasterEquationSchrodinger}) describes the average behaviour of the system, we briefly present the model describing stochastic evolution conditioned on the measurements results. In the case where a single output channel $\hat{L}$ is monitored by a photodetector, the stochastic master equation (or quantum filter) describing the dynamics is given by \cite{bkQMC}
\begin{equation}\label{eqQuantumFilter}
  d\hat{\rho}(t)  =\left(dt\mathcal{H}\left[-\frac{i}{\hbar}\hat{H}-\frac{\hat{L}^\dagger \hat{L}}{2}\right]+dN(t)\mathcal{G}[\hat{L}]\right)\hat{\rho}(t),
\end{equation}
where we define the superoperators
\begin{equation}\label{superop}
  \begin{aligned}
    \mathcal{H}[\hat{r}]\hat{\rho} &= \hat{r}\hat{\rho}+\hat{\rho}\hat{r}^\dagger-\mathrm{Tr}\left\{\hat{r}\hat{\rho}+\hat{\rho}\hat{r}^\dagger\right\}\hat{\rho}, \\
    \mathcal{G}[\hat{r}]\hat{\rho}&=\frac{\hat{r}\hat{\rho}\hat{r}^\dagger}{\mathrm{Tr}\left\{ \hat{r}\hat{\rho}\hat{r}^\dagger\right\}}-\hat{\rho}, 
  \end{aligned}
\end{equation}
with $\mathrm{Tr}\{\hat{r}\}$ denoting the trace of the operator $\hat{r}$.

Equation \cref{eqQuantumFilter} is an It\^o stochastic differential equation for the density matrix, which gives the increment to $\hat{\rho}$ in an infinitesimal time interval $dt$. The second term represents the action of a detection event on the state and is proportional to the stochastic increment $dN(t)$, which can take the values $0$ (no detection) or $1$ (detection). When a detection takes place, which happens with probability $\langle\hat{L}^\dagger\hat{L}\rangle(t) dt$ in the time interval $dt$, we say the system undergoes a `quantum jump': $\hat{\rho}$ is instanteneously replaced with $\hat{L}\hat{\rho}\hat{L}^\dagger$ and normalised, according to the definition of $\cal G$ in Eq. (\ref{superop}). The first term in Eq. \cref{eqQuantumFilter} corresponds to evolution in the lack of detection. From the definition of $\mathcal{H}$, this leads to the Hamiltonian unitary evolution $-\frac{i}{\hbar}[\hat{H},\hat{\rho}]$, as well as a conditioning term, $-\frac{1}{2}\hat{\rho}\hat{L}^\dagger\hat{L}-\frac{1}{2}\hat{L}^\dagger\hat{L}\hat{\rho} +\mathrm{Tr}\{\hat{L}^\dagger\hat{L}\hat{\rho}\}$. The conditioning occurs because a lack of clicks in the photodetector also provides us with information about the system, e.g. a long period without photodetections indicates that the cavity is likely empty. 

Note that if we average over many realizations of the stochastic trajectories described by Eq. \cref{eqQuantumFilter}, we recover the dynamics given by the master equation, Eq. \cref{eqMasterEquationSchrodinger}.


\section{Shaped single photons}\label{secSPG}

\begin{figure}[t!]
  \includegraphics[width=0.8\columnwidth]{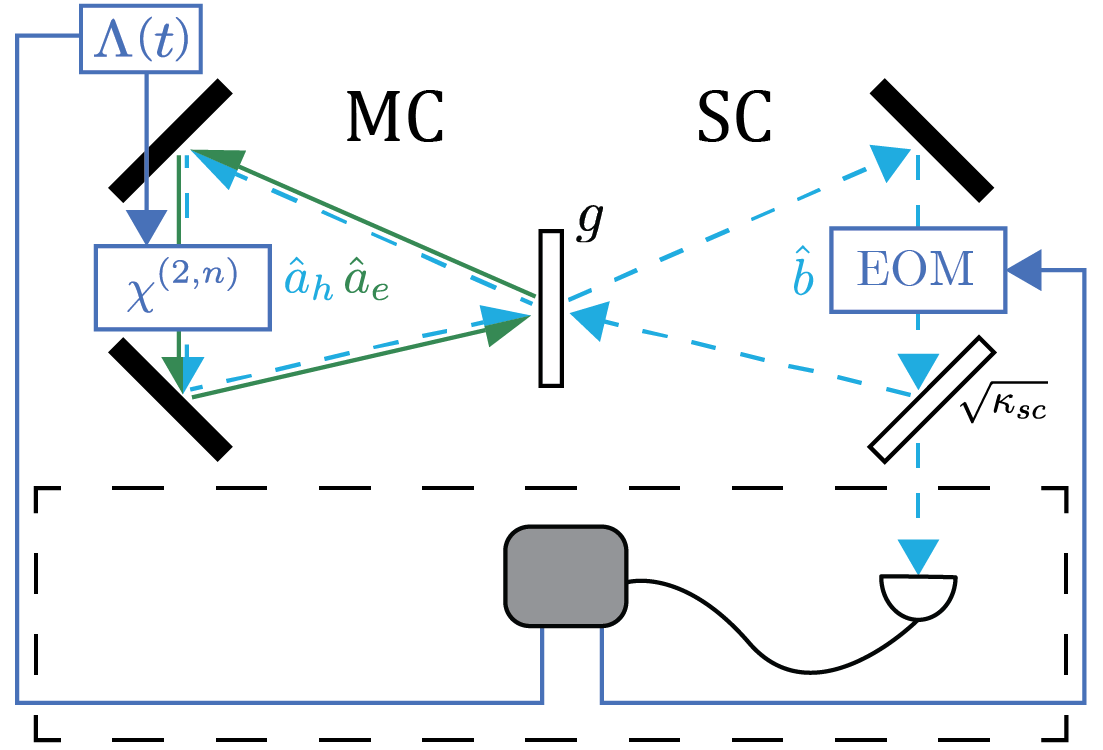}
  \caption{(Colour online) The scheme for shaped single photon generation consists of two cavities coupled by a semitransparent mirror. MC contains a $\chi^{(2,n)}$ optical nonlinearity, pumped by a coherent field $\Lambda(t)$. SC contains an EOM, which allows us to control its resonant frequency. MC is perfect while SC allows leakage at a rate $\kappa_{sc}$, with the output field continuously monitored by a photodetector.}\label{figSPG}
\end{figure}

To generate shaped, single photon states we use the scheme shown in~\cref{figSPG}, which is based on the experiment performed in~\cite{Yoshikawa:2013}. MC and SC are coupled via a semitransparent mirror with coupling constant $g$. MC is assumed perfect ($\kappa_e=\kappa_h=0$) and the nonlinearity it contains nondegenerate, so that the frequencies of the generated photons $\omega_h$ and $\omega_e$ are well separated. SC allows leakage at a rate $\kappa_{sc}$, with the output field ${\hat b}_{out}$ continuously monitored by a photodetector. 
We take $\Lambda(t)$ and $g$ to be real. 

The generation of shaped single photon occurs in two stages: `creation' and `readout'. 

\subsection{Creation}\label{secHeralding}

We first create a single $\hat{a}_e$ photon state in MC by waiting for a photodetection from SC. An intuitive picture for this follows from the experiment in~\cite{Yoshikawa:2013}. We tune $\omega_{sc}(t)=\omega_h$, and pump the nonlinearity until it generates a photon pair. Since the $\hat{a}_h$ mode is resonant with SC it will leak out and be detected. This `heralds' that we have a single $\hat{a}_e$ mode in MC, and we cease pumping before further pairs are created. The optimal regime will thus involve $\omega_{sc}(t)=\omega_h$, $g$, and $\kappa_{sc}$ large relative to $\Lambda(t)$ to lower the multiphoton components of the final state in MC, and $\kappa_{sc} \gg g$ so that the dynamics of the coupling do not play a significant role.

Since the detuning $|\omega_h-\omega_e|$ is large, we can disregard coupling of $\hat{a}_e$ into SC. As the coherent field $\Lambda(t)$ is constant during this stage, we will denote this as $\Lambda$. Moving to a frame rotating with respect to $\hbar\omega_h\hat{a}_h^\dagger\hat{a}_h+\hbar\omega_e\hat{a}_e^\dagger\hat{a}_e+\hbar\omega_h\hat{b}^\dagger\hat{b}$, the system is described by the Hamiltonian
\begin{equation}\label{eqSPGH}
  \hat{H}=\hbar\Lambda(\hat{a}_h\hat{a}_e+\hat{a}_h^\dagger\hat{a}_e^\dagger) + \hbar g(\hat{a}_h{{\hat b}}^\dagger+\hat{a}_h^\dagger {\hat b}),
\end{equation}
and loss operator
\begin{equation}\label{eqSPGL}
  \hat{L}=\sqrt{\kappa_{sc}} \, {\hat b}.
\end{equation}

The aforementioned parameter regime will cause the dynamics of ${\hat b}$ and $\hat{a}_h$ to be rapidly damped compared to those of $\hat{a}_e$, so we may simplify our picture by performing an adiabatic elimination~\cite{artAdiabaticEilimationUsed} of these modes, the details of which are provided in~\cref{appSPGAdiabaticElimination}. This shows that in the regime $g,\kappa_{sc}\gg\Lambda$, $\kappa_{sc}\gg g$, and $4g^2/\kappa_{sc}\gg\Lambda$, the system Eqs. \cref{eqSPGH} and \cref{eqSPGL} is equivalent to a single mode $\hat{a}_e$ evolving under a Hamiltonian and loss operator:
\begin{equation}\label{eqHeraldingParameters}
  \begin{aligned}
    \hat{H} &= 0,\\
    \hat{L} &= \sqrt{\gamma}\hat{a}_e^\dagger,
  \end{aligned}
\end{equation}
where 
\begin{equation}\label{eqSPGGamma}
  \gamma=\frac{\Lambda^2\kappa_{sc}}{g^2}.
\end{equation}
If we consider the evolution of Eq. \cref{eqHeraldingParameters} under Eq. \cref{eqQuantumFilter}, we see that a photodetection from the SC loss channel will indeed create a single emission mode in MC. Before this photodetection, the first term generates deterministic evolution $-dt\frac{\gamma}{2}\mathcal{H}[\hat{a}_e^\dagger\hat{a}_e]\hat{\rho}$ which, provided the system begins in the vacuum, will have no effect on the dynamics. However, this is only true in the ideal adiabatic limit. In any experimental implementation the dynamics of ${\hat b}$ and $\hat{a}_h$ are not instantaneous. The final state after photodetection will thus be a superposition of single and multiphoton components, which gets closer to a single photon as we approach the adiabatic limit. 

It may not be immediately clear why we require $4g^2/\kappa_{sc} \gg\Lambda$. With the adiabatic elimination it is derived as a sufficient rather than necessary constraint, however numerical simulations demonstrate that if we fix $g$ and $\Lambda$ then increase $\kappa_{sc}$ outside of this regime, the result is slower generation of the emission mode. This can be understood in terms of overdamped harmonic oscillation. If we consider the simple case with the nonlinearity unpumped ($\Lambda(t)=0$) and a photon pair $\hat{a}_h$, $\hat{a}_e$ in MC, the Heisenberg equation of motion for the heralding mode is
\begin{equation}
  \ddot{\hat{a}}_h+\frac{\kappa_{sc}}{2}\dot{\hat{a}}_h+g^2\hat{a}_h=0.
\end{equation}
We see that $\kappa_{sc}$ gives the damping rate of $\hat{a}_h$, and it is this damping that heralds the creation of $\hat{a}_e$. If $\kappa_{sc}$ grows too large we enter an overdamped regime, slowing the rate of production of $\hat{a}_e$.

After the output from SC has been detected, we cease pumping $\chi^{(2,n)}$, leaving the $\hat{a}_e$ mode stored in MC for later on-demand retrieval. Note that in our model we assumed no loss in the emission mode and therefore perfect storage, in reality the photon readout needs to happen within the cavity lifetime. 

\subsection{Readout}
To release the $\hat{a}_e$ mode stored in MC, we can tune $\omega_{sc}(t)=\omega_e$, allowing the photon to couple into SC and be emitted~\cite{Yoshikawa:2013}. Furthermore, we can control the strength of this interaction with the detuning $\Delta(t):=\omega_{sc}(t)-\omega_e$. To see this we consider the system in \cref{figSPG} with $\chi^{(2,n)}$ unpumped ($\Lambda(t)=0$), a single $\hat{a}_e$ mode in MC, and no $\hat{a}_h$ mode. In a frame rotating at $\omega_e\hat{a}_e^\dagger\hat{a}_e+\omega_e\hat{b}^\dagger\hat{b}$, the system Hamiltonian and decoherence operators are then
\begin{equation}
  \begin{aligned}
    H&=\hbar\Delta(t)\hat{b}^\dagger\hat{b}+\hbar g\left(\hat{a}_e\hat{b}^\dagger+\hat{a}_e^\dagger\hat{b}\right),\\
    L&=\sqrt{\kappa_{sc}}\hat{b}.
  \end{aligned}
\end{equation}
To simplify the picture we adiabatically eliminate the strongly damped $\hat{b}$ mode using the same method as in \cref{appSPGAdiabaticElimination}. The result is a system in terms of the $\hat{a}_e$ mode only, which has Hamiltonian and decoherence operator
\begin{equation}\label{eqEmissionHL}
  \begin{aligned}
    \hat{H} &= 2\hbar g^2\Delta(t)\hat{a}_e^\dagger\hat{a}_e, \\
    \hat{L} &= \sqrt{\frac{2g^2\kappa_{sc}}{4\Delta(t)^2+\kappa_{sc}^2}}\hat{a}_e.
  \end{aligned}
\end{equation}
The only assumption required in deriving \cref{eqEmissionHL} is that $\kappa_{sc}^2\gg g^2$, which follows from the $\kappa_{sc}\gg g$ needed in Section \cref{secHeralding}. The Hamiltonian leads to an oscillation in phase which may be neglected for our purposes, and so we see that the net effect of SC in this regime is to provide a loss channel for $\hat{a}_e$ whose strength depends on the detuning $\Delta(t)$.

Since the output photon is not continuously monitored but rather used as input to some other system, we are interested in the behaviour described by the master equation Eq. \cref{eqMasterEquationSchrodinger}:
\begin{equation}\label{eqEmissionAEEOM}
  \frac{d}{dt}\hat{\rho}_e=-2ig^2\Delta(t)[\hat{a}_e^\dagger \hat{a}_e,\hat{\rho}_e]+\frac{2g^2\kappa_{sc}}{4\Delta(t)^2+\kappa_{sc}^2}\mathcal{D}[\hat{a}_e]\hat{\rho}_e.
\end{equation}
For a decaying system $\rho$ obeying the equation of motion
\begin{equation}\label{eqDecayingSystem}
  \frac{d}{dt}\hat{\rho}=\lambda(t)\mathcal{D}[\hat{a}]\hat{\rho},
\end{equation}
if the coefficient $\lambda(t)$ can be controlled arbitrarily, then we can generate any desired output shape~\cite{artFilteringTwoLevel}. To achieve this we define a function $\xi(t)$ such that $\lvert\xi(t)\rvert^2$ gives the desired temporal profile, with normalisation $\int_{-\infty}^\infty\lvert\xi(t)^2\rvert dt=1$. For the output of the cavity to match this, we require 
\begin{equation}\label{eqLambdaForShaping}
  \lambda(t)=\frac{\xi(t)}{\sqrt{\int_t^\infty\lvert \xi(t')^2\rvert dt'}}.
\end{equation}
Our system in Eq. \cref{eqEmissionAEEOM} approximates that of Eq. \cref{eqDecayingSystem}, with $2g^2\kappa_{sc}/(4\Delta(t)^2+\kappa_{sc}^2)$ playing the role of $\lambda(t)$. There are however two differences, the first being that while we can make the loss rate arbitrarily small by increasing $\Delta(t)$, it cannot grow larger than $\frac{2g^2}{\kappa_{sc}}$, which occurs when $\Delta(t)=0$. To approximate some desired $\lambda(t)$ we thus choose
\begin{equation}\label{eqDeltaForShaping}
  \Delta(t)=
  \begin{cases}
    \frac{1}{2}\sqrt{\frac{2g^2\kappa_{sc}}{\lambda(t)} -\kappa_{sc}^2} & \lambda(t) < \frac{2g^2}{\kappa_{sc}}, \\
    0 & \lambda(t) \ge\frac{2g^2}{\kappa_{sc}},
  \end{cases}
\end{equation}
with the output pulse shape growing closer to $\lvert\xi(t)\rvert^2$ as the ratio $\frac{2g^2}{\kappa_{sc}}$ increases.

\begin{figure}[t!]
  \center
  \includegraphics[width=0.8\columnwidth]{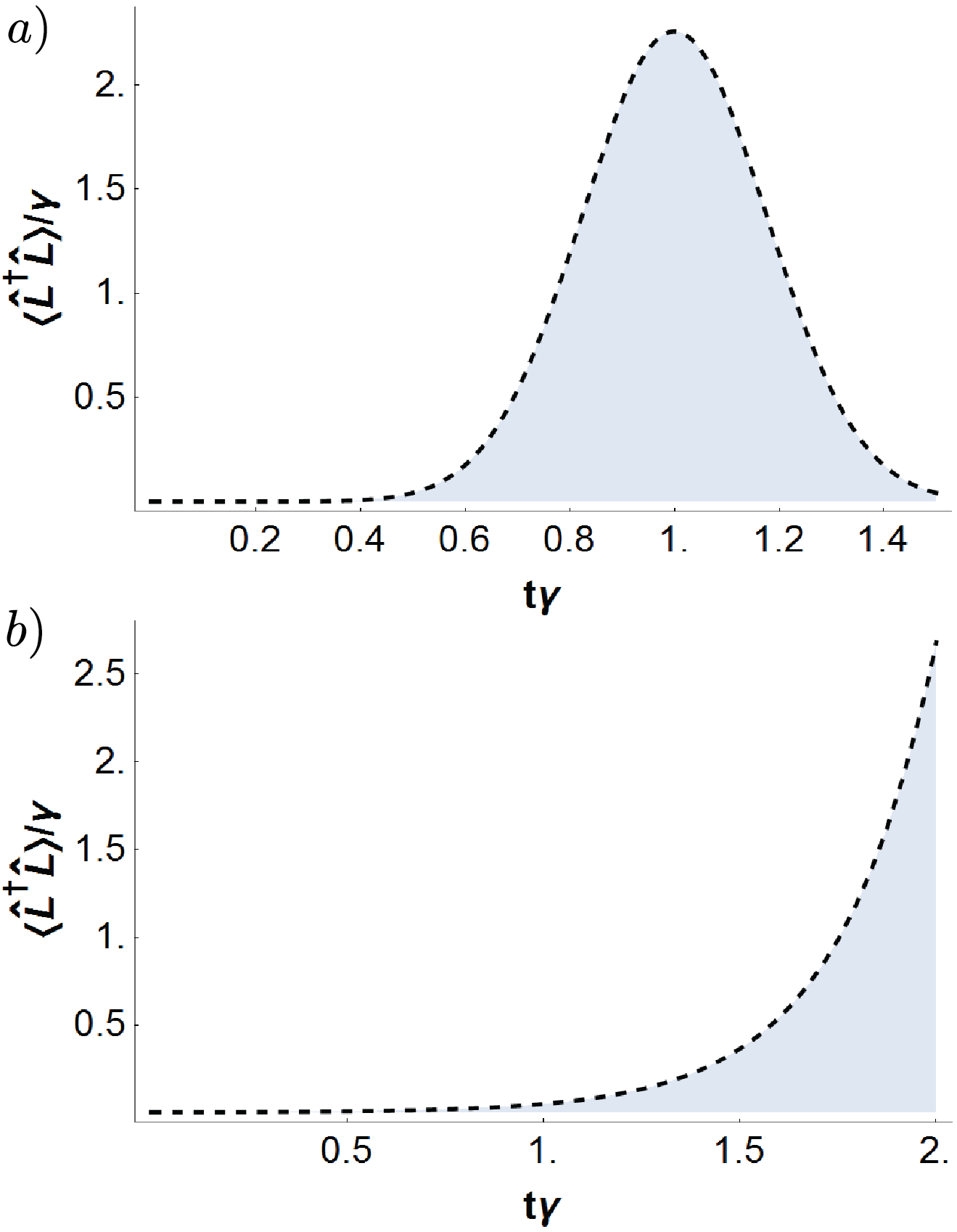}
  \caption{Simulations of Eq. \cref{eqEmissionAEEOM} during readout stage, with $\Delta(t)$ chosen according to Eq. \cref{eqDeltaForShaping} in order to generate a) Gaussian and b) rising exponential pulse shapes. The dashed line shows the desired temporal profile $\lvert\xi(t)\rvert^2$, while the solid region is the emission probability density Eq. \cref{eqEmissionProbability}. We begin with a single photon in the $\hat{a}_e$ mode, and choose parameters $g=100\gamma$ and $\kappa_{sc}=10g$ (with $\Lambda=0$ during readout).}\label{figPulseShaping}
\end{figure}

The other difference is the time-dependent phase term $-2ig^2\Delta(t)a_e^\dagger a_e$ which is present in Eq. \cref{eqEmissionAEEOM} but not Eq. \cref{eqDecayingSystem}. This simply rotates the phase of the output field, and should not have an effect on the pulse shape (though it may need to be taken into account if the generated pulse will be used in interference experiments). This is confirmed by numerical simulations \cite{artQuTiP,artQuTiP2,artXMDS} in \cref{figPulseShaping}, which show Gaussian and rising exponential pulse shapes generated by choosing $\Delta(t)$ according to Eq. \cref{eqDeltaForShaping}. Using Eq. \cref{eqEmissionHL}, the emission probability density of a photon from the system is given by 
\begin{equation}\label{eqEmissionProbability}
  \langle\hat{L}^\dagger\hat{L}\rangle(t)=\frac{2g^2\kappa_{sc}}{4\Delta(t)^2+\kappa_{sc}^2}\langle\hat{a}_e^\dagger\hat{a}_e\rangle(t).
\end{equation}

\section{Generation of coherent-state superpositions}\label{secAnotherChannel}

\begin{figure}[t!]
  \includegraphics[width=0.9\columnwidth]{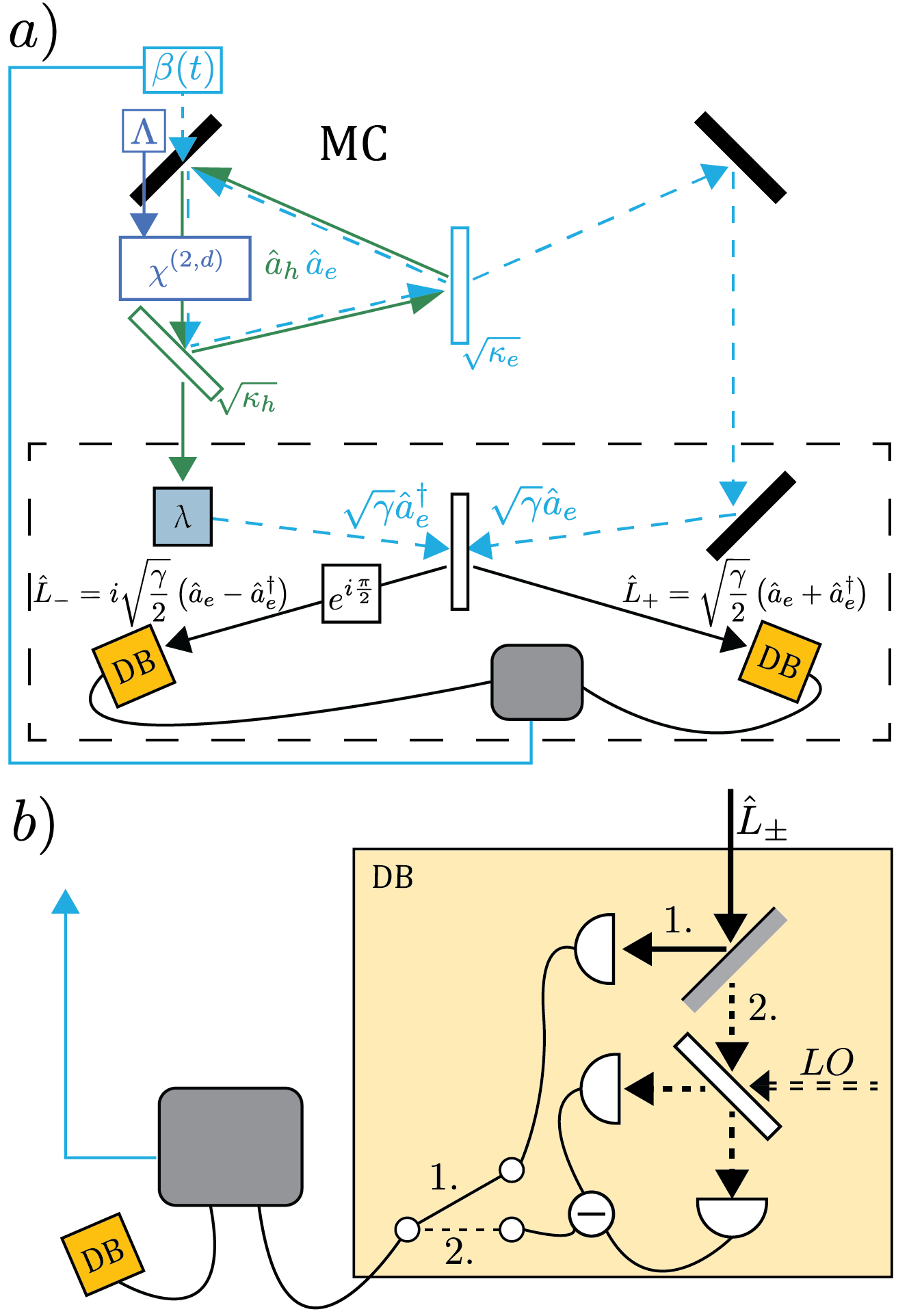}
  \caption{(Colour online) a) The system for generation of CSS consists of a single cavity, MC, containing a $\chi^{(2,d)}$ optical nonlinearity, pumped by a coherent field of constant amplitude $\Lambda$. Polarisation-sensitive semitransparent mirrors control which mode is emitted from each mirror, and a waveplate $\lambda$ matches the polarisations so that the two fields are indistinguishable. They are interfered through a 50/50 beamsplitter, then have a relative phase of $\frac{\pi}{2}$ introduced leading to two Hermitian decoherence channels which are measured by detection blocks DB. During the no-knowledge feedback stage, a coherent field with frequency $\omega_e$ and complex amplitude $\beta(t)$ proportional to the measurement signal is introduced into the cavity. b) A schematic of each DB, which allows us to choose between photodetection and homodyne measurement. During state generation (1.) the decoherence channel is directed along the path indicated by the solid arrows to the photodetector, where the grey block denotes a mirror of controllable reflectivity, and a simple switch in the bottom left controls which signal is received by the controller. For storage via No-Knowledge Feedback, the mirror is made transmissive and the channel follows the dashed arrows towards a homodyne measurement (2.), where it is interfered with a local oscillator ($LO$) and the two beamsplitter outputs are subtracted. The corresponding photocurrent is used to modulate $\beta(t)$ incident on MC, which has the effect of removing the decoherence channel from the system dynamics \cite{Szigeti:2014}.}\label{figSystemExtraChannel}
\end{figure}

Figure~\ref{figSystemExtraChannel} a) shows the changes in the setup needed to generate coherent-state superpositions. The nonlinearity is now degenerate, creating pairs of photons $\hat{a}_e$, $\hat{a}_h$ at the same frequency $\omega$ but orthogonal polarisations, indicated by the dashed and solid lines. The pump field has constant amplitude $\Lambda$, which we take to be real. We have a single cavity, MC, and polarisation sensitive mirrors cause the heralding and emission modes to be emitted from separate cavity outputs at rates $\kappa_h$ and $\kappa_e$ respectively. We also allow for a coherent field $\beta(t)$ to be introduced to the cavity. This is zero during state preparation, and we will later use it to perform feedback during the storage phase. As in the previous section we choose parameters so that the dynamics of $\hat{a}_h$ are short lived, and adiabatically eliminate this mode. Defining 
\begin{equation}\label{eqCSSGamma}
  \gamma=\frac{4\Lambda^2}{\kappa_h}
\end{equation}
and choosing $\kappa_{e}=\gamma$, the loss operators $\hat{L}_i$ corresponding to output immediately after the $\kappa_i$ mirror become
\begin{equation}\label{eqLMCSC}
  \begin{aligned}
    \hat{L}_{e}&=\sqrt{\gamma}\hat{a}_e, \\
    \hat{L}_{h}&=\sqrt{\gamma}\hat{a}_e^\dagger.
  \end{aligned}
\end{equation}

We can intuitively understand Eq. \cref{eqLMCSC} by considering what information a photodetection from each cavity would provide. Detection of a photon from $\kappa_e$ indicates loss of an emission mode, hence $\hat{L}_{e}\propto\hat{a}_e$. A photodetection from $\kappa_h$ indicates that a heralding mode has been emitted. As the modes are created in pairs, this informs us that there must be a corresponding emission mode in MC, and so $\hat{L}_{h}\propto\hat{a}_e^\dagger$ in the eliminated regime.

The two cavity outputs differ only by polarisation. A waveplate is used to render them indistinguishable, and they are interfered through a beamsplitter. If only heralded CSS state generation is required, the two resulting channels may be monitored by photodetectors. However, later we will discuss storage of the CSS state with No-Knowledge Feedback (NKF) \cite{Szigeti:2014}. NKF requires Hermitian decoherence channels and homodyne measurement. Thus we introduce a relative phase shift of $\frac{\pi}{2}$, and then direct the outputs towards measurement blocks (\cref{figSystemExtraChannel} b), which allow for switching to a homodyne measurement after the required state has been prepared.

Let us first consider state generation. If we were to monitor the channels Eq. \cref{eqLMCSC} with photodetectors placed immediately after the semitransparent mirrors, a detection from the $\hat{L}_{j}$ channel, where $j\in\{h,e\}$, would correspond to acting $\hat{L}_{j}$ on the state in MC. Beginning from the vacuum, these operators would prepare $n$-photon states. In order to generate CSS we interfere the channels as shown in \cref{figSystemExtraChannel} a), resulting in
\begin{equation}
  \begin{aligned}
    \hat{L}_+ &= \sqrt{\frac{\gamma}{2}}\left(\hat{a}_e+\hat{a}_e^\dagger \right), \\
    \hat{L}_- &= i\sqrt{\frac{\gamma}{2}}\left(\hat{a}_e-\hat{a}_e^\dagger \right).
  \end{aligned}
\end{equation}
We can recognise these in terms of the quadrature operators: 
\begin{equation}
  \begin{aligned}
    \hat{x} &= \frac{1}{\sqrt{2}}\left(\hat{a}+\hat{a}^\dagger\right), \\
    \hat{p} &= \frac{i}{\sqrt{2}}\left(\hat{a}^\dagger-\hat{a}\right).
  \end{aligned}
\end{equation}
Direct photodetection of the $\hat{L}_{\pm}$ corresponds to the instantaneous action of the corresponding operator on the system, thus leading to quantum jumps proportional to field quadratures~\cite{Santos:2011}. 

Let us suppose we begin from the vacuum and have $n$ photodetections from the $\hat{L}_+$ channel in quick succession, so that we may neglect the evolution of the system in-between jumps (the effect of this will be considered later). We will show that this produces, to a good approximation, a superposition of coherent states in MC. The cavity state after $n$ photodetections from the $\hat{L}_+$ channel will be
\begin{equation}\label{eqLMinusRepeatedHOBasis}
  |\tilde{\psi}_n\rangle=\hat{L}_+^n|0\rangle=\gamma^{\frac{n}{2}}\hat{x}^n|0\rangle,
\end{equation}
where the tilde represents unnormalised states. 
The normalised version $\ket{\psi_n}$ in the position basis can be written as 
\begin{equation}\label{eqLMinusRepeatedPositionBasis}
  \ket{\psi_n}=\frac{1}{\sqrt{\Gamma\left(n+\frac{1}{2}\right)}}\int_{-\infty}^{\infty} dx x^ne^{-\frac{x^2}{2}}|x\rangle,
\end{equation}
where $\Gamma(z)$ is the Euler gamma function. Note that the wavefunction  $x^ne^{-\frac{x^2}{2}}$ is symmetric(anti-symmetric) about $x=0$ if $n$ is even(odd), with peaks at
\begin{equation}\label{eqLnZeroWavefunctionPeaks}
  x=\pm\sqrt{n}.
\end{equation}

\begin{figure}[t!]
    \centering
    \includegraphics[width=0.9\columnwidth]{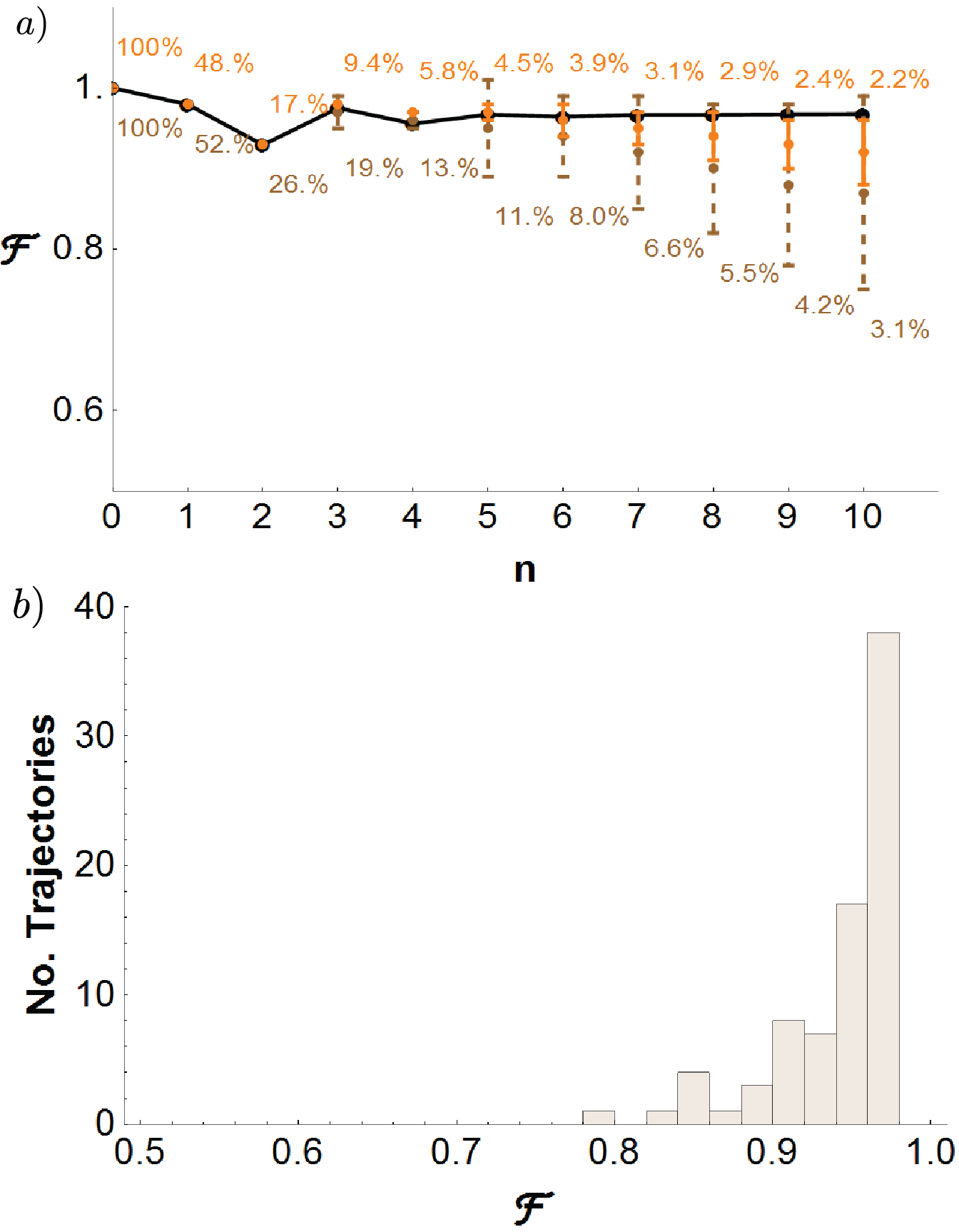}
    \caption{(Colour online) We plot in a) the fidelity $\mathcal{F}$ of the state generated by $n$ consecutive detections from the $\hat{L}_+$ channel with the normalised version of the CSS in Eq. \cref{eqAlphaAnsatz}. The solid (black) line shows the ideal fidelity given by \cref{eqFidelity}, if the jumps occurred instantaneously with negligible evolution in-between. We see that this rapidly approaches a value of around $0.97$. The effect of the `no-jump evolution' is shown by the vertical series, which simulate 1,000 trajectories for a time $t\gamma=5$ and plot the average fidelity of the postselected states with Eq. \cref{eqAlphaAnsatz}. The error is taken to be the variance, and indicated by the range of the brackets. The numbers accompanying each series give the percentage of trajectories that survived postselection, which we interpret as the probability of the state being generated. The dashed (brown) series denotes postselection only for the correct jump sequence, while the solid (orange) series also selects for a maximum time between jumps of $t\gamma=\frac{1}{2}$. In b) we show a histogram of the fidelities of the ideal state Eq. \cref{eqAlphaAnsatz} with the generated CSS states for $n=6$, where we postselect only on jump sequence. We see that this is concentrated at the ideal value of $0.97$.}
    \label{figFidelityOfCats}
\end{figure}

To show that $\ket{\psi_n}$ is approximately a Schr\" odinger cat state, we consider an ansatz where the peaks of the CSS coincide with Eq.~\cref{eqLnZeroWavefunctionPeaks}. 
For a coherent state  $|\alpha\rangle$, the peak in the position basis is located at 
\begin{equation}\label{eqCSWavefunctionPeak}
  x=\sqrt{2}\alpha,
\end{equation}
where we assume $\alpha$ to be real. Equating \cref{eqLnZeroWavefunctionPeaks} and \cref{eqCSWavefunctionPeak}, we make the ansatz that $|\psi_n\rangle$ is approximated by the superposition of coherent states:
\begin{equation}\label{eqAlphaAnsatz}
 \ket{\tilde \phi_n}= \left\lvert\sqrt{\frac{n}{2}}\right\rangle+(-1)^n\left\lvert-\sqrt{\frac{n}{2}}\right\rangle.
\end{equation}
The fidelity $\mathcal {F}=\vert \langle \psi_n \ket{\phi_n} \vert$ between our generated state Eq. \cref{eqLMinusRepeatedHOBasis} and the CSS \cref{eqAlphaAnsatz} can be readily evaluated to be 
\begin{equation}\label{eqFidelity}
 \mathcal{F} = 
 \begin{cases}
 \frac{\Gamma\left(\frac{n+1}{2}\right){}_1F_1\left(-\frac{n}{2},\frac{1}{2},-\frac{n}{4}\right)}{\pi^{\frac{1}{4}}\sqrt{\mathrm{cosh}\left(\frac{n}{2}\right)\Gamma\left(n+\frac{1}{2}\right)}} & n\;\mathrm{even}, \\
 \\
 \frac{\sqrt{n}\Gamma\left(\frac{n}{2}+1\right){}_1F_1\left(-\frac{n-1}{2},\frac{3}{2},-\frac{n}{4}\right)}{\pi^{\frac{1}{4}}\sqrt{\mathrm{sinh}\left(\frac{n}{2}\right)\Gamma\left(n+\frac{1}{2}\right)}} & n\;\mathrm{odd},
 \end{cases}
\end{equation}
where ${_1}F_1(a,b,c)$ is Kummer's confluent hypergeometric function. In~\cref{figFidelityOfCats} a the solid (black) line plots the fidelities of our generated state with the ansatz coherent state superposition for $n\le10$. This rapidly approaches a constant value of approximately $0.97$. Finally we note that if we instead consider $\hat{L}_-^n|0\rangle$ we arrive at a similar result, with $\alpha$ for the coherent superposition now lying along the imaginary axis. 

\begin{figure*}[!bt]
  \includegraphics[width=\textwidth]{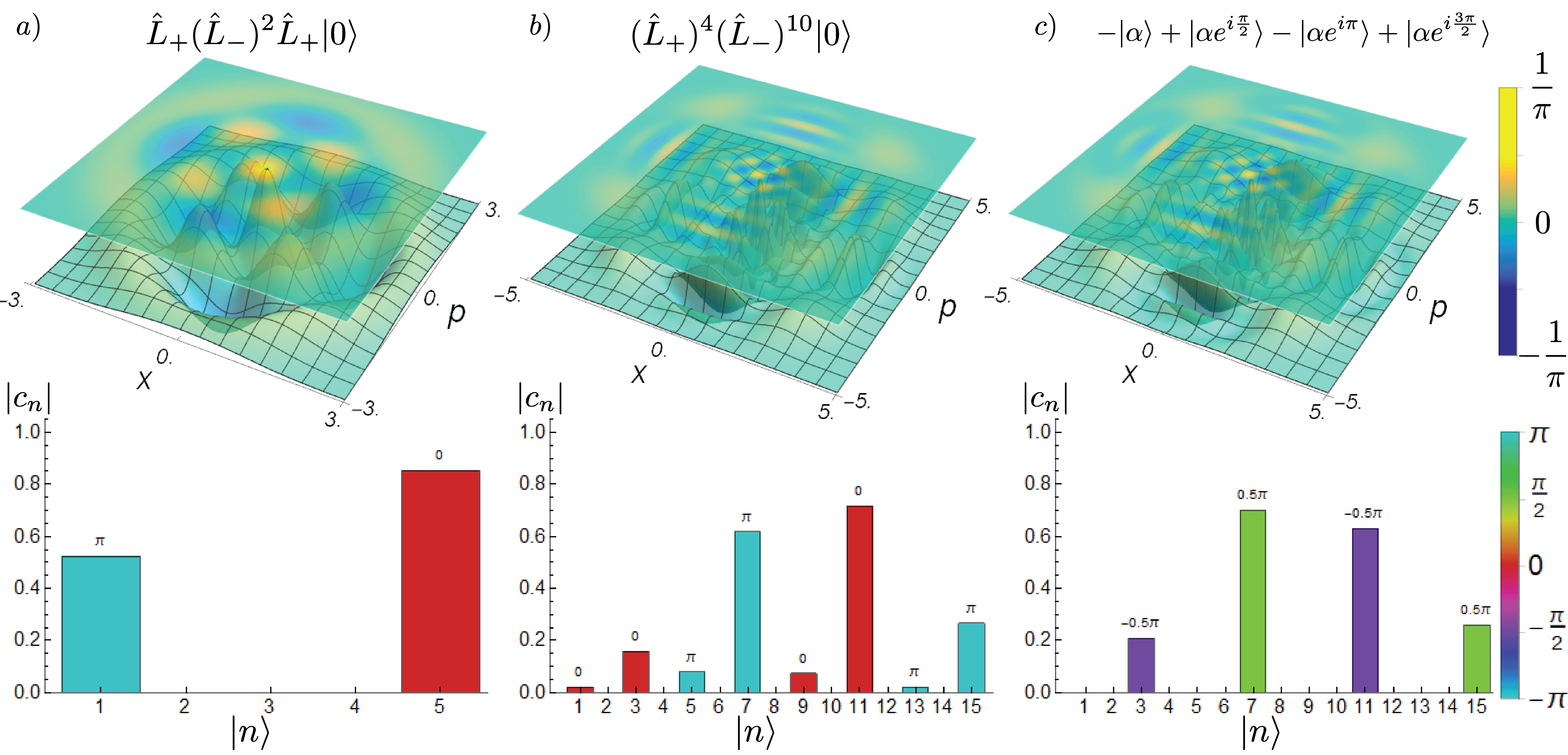}
  \caption{We show Wigner functions and Fock basis expansions for a) a superposition of the Fock states $|1\rangle$ and $|4\rangle$ and b) a CSS which approximates a four-component coherent state superposition, both generated through sequences of detections from the scheme in \cref{figSystemExtraChannel}. In c) we shown an actual ``Schr\"odinger's cat' coherent states superposition (the Fock basis expansion is shown up to $n=14$), where $\alpha=2.825e^{i\frac{\pi}{4}}$. The CSS approximates this state with a fidelity of $0.98$. In the upper plots colour denotes the height of the Wigner function. For the lower plots colour and the series labels denote the complex phase of the Fock basis coefficients.}\label{figFockSuperpositions}
\end{figure*}

Other sequences of jumps can also prepare interesting states. We show in~\cref{figFockSuperpositions} two examples of this: a) a superposition of Fock states and b) a state approximating a four-component cat state. While the latter is shown to require a sequence of fourteen jumps, different sequences can generate similar states (often with rotated phase), and we can generate a lower amplitude state with fewer jumps.

So far there are two factors that we have neglected in our discussion. The first is the stochastic nature of the detections. Jumps occur randomly from either channel, and so to observe a particular sequence we must postselect for favourable trajectories. For example, if we wish to generate the state $\hat{L}_+^3\lvert0\rangle$, we begin with an empty cavity and wait for three successive detections from the $\hat{L}_+$ channel. Before this happens, if we register a photodetection from $\hat{L}_-$, we empty the cavity and begin again. 

The efficiency of this solution decreases for large $n$. This is shown in \cref{figFidelityOfCats} a by the lower (brown) series of percentages, which denote the fraction of 1000 simulated trajectories that registered $n$ successive jumps from $\hat{L}_+$. We make the observation that the percentages are greater than $2^{-n}$, which one might naively expect for two jump channels with equal probability. For example, the state $\hat{L}_+^{10}|0\rangle$ was generated $3.1\%$ of the time, which is much larger than $2^{-10}\approx 0.1\%$. This is because the jumps do not have equal probability. The probability of a click from the $\hat{L}_+$ channel in a time interval $dt$ is given by the expectation value $\langle\hat{L}^\dagger_+\hat{L}_+\rangle dt$, which will generally be different from $\langle\hat{L}^\dagger_-\hat{L}_-\rangle dt$. 

The second factor is the dynamics between jumps, which will reduce fidelity with the target state. To understand this we can look at the equation of motion for the state of the cavity in the absence of jumps. By considering only the deterministic first term of Eq. \cref{eqQuantumFilter}, now summing over the $\hat{L}_{\pm}$, we find that in the absence of photodetections the state evolves as 
\begin{equation}\label{eqFockSuperpositionsNoJumpEvolution}
  \begin{aligned}
    \frac{d\hat{\rho}}{dt}&=-\sum_{j\in\{+,-\}}\mathcal{H}\left[\frac{\hat{L}_j^{\dagger}\hat{L}_j}{2}\right]\hat{\rho},\\
     &=\gamma\left(2\langle\hat{n}\rangle\hat{\rho}-\hat{n}\hat{\rho}-\hat{\rho}\hat{n}\right),
  \end{aligned}
\end{equation}
where $\hat{n}=\hat{a}_e^\dagger \hat{a}_e$ is the number operator for the cavity mode. Over time this will drive a superposition towards the Fock state with the smallest $n$ present at a rate proportional to $\gamma$. The probability of a photodetection in the time interval $dt$ is also proportional to $\gamma$:
\begin{equation}
  \langle L_j^\dagger L_j\rangle dt\propto\gamma dt,
  \end{equation}
so we cannot mitigate the effect of Eq. \cref{eqFockSuperpositionsNoJumpEvolution} by increasing the rate at which photodetections occur.

The influence of the no-jump evolution on the fidelity of the generated states is shown by the dashed (brown) vertical series in \cref{figFidelityOfCats} a. This plots the mean fidelity of states generated by postselection on 1000 simulated trajectories with the ideal state in \cref{eqFidelity}, with the error taken to be the variance. We see that the no-jump evolution does lead to a reduction in fidelity, however as seen in \cref{figFidelityOfCats} b the distribution is concentrated on the ideal value. Furthermore, the solid (orange) vertical series shows that by selecting for successive jump times less than $t\gamma=\frac{1}{2}$, we can create states with a very high average fidelity without significantly reducing the percentage of trajectories which survive postselection.

While mostly detrimental, this no-jump evolution can also be used to improve the generation of particular states. Take for example the Fock state superposition in \cref{figFockSuperpositions} a. If one wished to move from this state to an equal superposition $\frac{1}{\sqrt{2}}(|0\rangle+|4\rangle)$, we could postselect on the lack of jumps over a time interval to balance the coefficients of the $\ket{0}$ and $\ket{4}$ components.

Once the desired state has been prepared, for heralded release we cease pumping $\chi^{(2,d)}$, and redirect the output from $\kappa_e$ for readout in the desired direction. Alternatively, we can store the state in MC for on-demand release at a later time by use of a a recently proposed feedback technique to suppress decoherence~\cite{Szigeti:2014}. This `No-Knowledge Feedback' (NKF) allows us to cancel the effect of any Hermitian decoherence channel $\hat{L}$ by performing a homodyne measurement at an angle of $\frac{\pi}{2}$, as shown in \cref{figSystemExtraChannel} b). Such a measurement yields no information about the system, however the system dynamics under this continuous measurement become unitary, generated by an effective Hamiltonian 
\begin{equation}
  \hat{H}_{\mathrm{eff}}=\hat{H}-\hat{L}j(t),
\end{equation}
where $j(t)$ is the homodyne photocurrent, with no effective decoherence channel. The effect of this can be cancelled by feeding back the measured signal into the system, modifying the Hamiltonian by
\begin{equation}
  \hat{H}\rightarrow \hat{H}+\hat{L}j(t).
\end{equation}
We may apply this principle to individually cancel out both the Hermitian $\hat{L}_{\pm}$ channels. This corresponds to feeding back terms of the form $\hat{a}+\hat{a}^\dagger$ and $i(\hat{a}-\hat{a}^\dagger)$ into MC, which can be achieved by a coherent field into the cavity, as indicated by $\beta(t)$ in Fig~\ref{figSystemExtraChannel}.

This ability to freeze decoherence channels at will can also be used to prepare different kinds of states. For example, if we apply NKF to the $L_-$ channel, then only $L_+$ jumps will occur and the dynamics between jumps changes from Eq. \cref{eqFockSuperpositionsNoJumpEvolution} to 
\begin{equation}\label{eqNKFNoJump}
  \frac{d\hat{\rho}}{dt}=\frac{\gamma}{2}\left(2\langle\hat{x}^2\rangle\hat{\rho}-\hat{x}^2\hat{\rho}-\hat{\rho}\hat{x}^2\right).
\end{equation}

Note that even though the sequence of jumps is now completely deterministic, CSS are not produced in this case due to the change in the no-jump term. In fact, numerical simulations show that this new dynamics rapidly generates squeezed states. Other useful states could be generated with NKF, which may be considered in future work.

No Knowledge Feedback is not the only way in which storage of the CSS may be achieved. Instead of the mirror $\kappa_e$ leading directly to the detection scheme, we could adopt a method similar to the scheme in III and place a `shutter cavity' SC in-between containing an EOM, which is initially made resonant with MC, and whose parameters are chosen so that the SC mode may be adiabatically eliminated. Once the CSS has been created we cease pumping the $\chi^{(2,d)}$ and detune SC from MC, trapping the $\hat{a}_e$ mode until the state is required.

\section{Conclusion}
We have outlined a state generation scheme that can be used for on-demand production of various non-classical states of light. It allows for the production of on-demand shaped single photon states, with very little modification from the already implemented scheme from~\cite{Yoshikawa:2013}. We also showed how to generate coherent-state superpositions by exploring different detection strategies and using postselection. The CSS state can be stored in the cavity using a No-Knowledge Feedback scheme for later on-demand release. The main limitation to this is decoherence effects between detections, which can be minimised by another layer of postselection based on time between detection events.

The scheme outlined in \cref{secAnotherChannel} can be expanded in many ways. Using combinations of beamsplitters and phase shifts, we could manufacture other loss channels than the $\hat{L}_{\pm}$, allowing for generation of a wider class of states. Implementing the phase shifts via an EOM could allow for loss channels which change dynamically depending on the sequence of jumps which have occurred. States which are reasonably robust against the evolution in Eq. (\ref{eqNKFNoJump}) could also be prepared deterministically using No-Knowledge Feedback.

Given the similarity with the optical setup in \cite{Yoshikawa:2013}, both schemes can be implemented in an experiment. For on-demand shaped single photon generation, the primary challenge is to store the state for a time period comparable to the rate of pair production in the nonlinearity. In \cite{Yoshikawa:2013} the cavity storage time was around $1\mu s$ while pair production occurred on average every $3ms$, however the authors noted much potential for closing this gap. For a NKF scheme when preparing CSS states, the primary limitation would be implementation of the feedback on the system with minimal noise and time delay.

\section{Acknowledgements}

We would like to thank A. Furusawa, G. Zhang, N. Yamamoto, G Campbell, and R. Taylor for enlightening discussions. We gratefully acknowledge support  by the  Australian Research Council Centre of Excellence for Quantum Computation and Communication Technology (project number CE110001027), and an Australian Government Research Training Program (RTP) Scholarship. M.R.H. acknowledges funding from an Australian Research Council (ARC) Discovery Project (Project No. DP140101779).

\appendix

\section{Adiabatic Elimination in \cref{secHeralding}}\label{appSPGAdiabaticElimination}
In this section we provide an outline of the adiabatic elimination performed in \cref{secHeralding}, which follows the method outlined in \cite{artAdiabaticEilimationUsed}. The system is described in a rotating frame by a Hamiltonian $\hat{H}$ and loss operator $\hat{L}$:
\begin{equation}
  \begin{aligned}
    \hat{H} &=\hbar\Lambda(\hat{a}_h\hat{a}_e+\hat{a}_h^\dagger\hat{a}_e^\dagger) + \hbar g(\hat{a}_h\hat{b}^\dagger+\hat{a}_h^\dagger \hat{b}), \\
    \hat{L}&=\sqrt{\kappa_{sc}}\hat{b},
  \end{aligned}
\end{equation}
with the assumption $g,\kappa_{sc}\gg\Lambda$ and $\kappa_{sc}\gg g$.
We perform this calculation calculation using the master equation for an unmonitored system (Eq. \cref{eqMasterEquationSchrodinger}) with a single decoherence operator:
\begin{equation}\label{eqMasterEquationSchrodingerApp}
  \frac{d}{dt}\hat{\rho}=-\frac{i}{\hbar}[\hat{H},\hat{\rho}]+\mathcal{D}[\hat{L}]\hat{\rho},
\end{equation}
from which we can later extract $\hat{H}$ and $\hat{L}$.

We begin by partially expanding the density matrix for the entire system, $\hat{\rho}_{MC\otimes SC}$, over the Hilbert space corresponding to $\hat{b}$:

\begin{equation}\label{eqRhoFullSystemExpansion}
  \begin{aligned}
    \hat{\rho}_{MC\otimes SC}=&\hat{\rho}_{00}|0\rangle\langle 0|+\hat{\rho}_{10}|1\rangle\langle 0|+\hat{\rho}_{01}|0\rangle\langle 1|\\
                             &+\hat{\rho}_{20}|2\rangle\langle 0|+\hat{\rho}_{02}|0\rangle\langle 2| + o(\zeta^3),
  \end{aligned}
\end{equation}
where 
\begin{equation}
  \zeta=\frac{g}{\kappa_{sc}},
\end{equation}
kets $|i\rangle$ exist in the Hilbert space $\mathcal{H}_{\hat{b}}$ of $\hat{b}$, and $\hat{\rho}_{ij}$ acts on $\mathcal{H}_{MC}:=\mathcal{H}_{\hat{a}_h\otimes \hat{a}_e}$.

We substitute Eq. \cref{eqRhoFullSystemExpansion} into the master equation Eq. \cref{eqMasterEquationSchrodingerApp}, which we use to find equations of motion for the $\hat{\rho}_{ij}$ by acting $\langle i|\cdot|j\rangle$. To the first order in $\zeta$ we derive
\begin{equation}\label{eqRhoEliminationBR20DE}
  \frac{d}{dt}\hat{\rho}_{20}= -i\sqrt{2}g\hat{a}_h\hat{\rho}_{10}-\kappa_{sc}\hat{\rho}_{20}+o(\zeta^2).
\end{equation}
The assumption that $\kappa_{sc}\gg g$ corresponds to $\frac{d}{dt}\hat{\rho}_{20}\approx 0$, which can be justified mathematically by considering  the solution to Eq. \cref{eqRhoEliminationBR20DE}, $\hat{\rho}_{20}(t)=-i\sqrt{2}g\int_0^te^{-\kappa_{sc}(t-t')}\hat{a}_h\hat{\rho}_{10}(t')dt'$, in the limit $\kappa_{sc}\gg g$. With this Eq. \cref{eqRhoEliminationBR20DE} yields
\begin{equation}\label{eqRhoEliminationBR20}
  \hat{\rho}_{20}\approx -i\sqrt{2}\frac{g}{\kappa_{sc}}\hat{\rho}_{10}.
\end{equation}
Using Eq. \cref{eqRhoEliminationBR20} and proceeding similarly for $\hat{\rho}_{10}$ we find
\begin{equation}\label{eqRhoEliminationBR10}
  \hat{\rho}_{10}=-2i\frac{g}{\kappa_{sc}}(\hat{a}_h\hat{\rho}_{00}-\hat{\rho}_{11}\hat{a}_h).
\end{equation}
We can use Eq. \cref{eqRhoEliminationBR10} to find an expression for $\hat{\rho}_{MC}$, the density operator over $\mathcal{H}_{MC}$. As $\frac{d}{dt}\hat{\rho}_{10}=\frac{d}{dt}\hat{\rho}_{20}=0$:
\begin{equation}\label{eqSPGAhEliminated}
  \begin{aligned}
  \frac{d}{dt}\hat{\rho}_{MC}&=\frac{d}{dt}\left(\hat{\rho}_{00}+\hat{\rho}_{11}\right), \\&=-i[\Lambda(\hat{a}_e\hat{a}_h+\hat{a}_e^\dagger \hat{a}_h^\dagger),\hat{\rho}_{MC}]+\frac{4g^2}{\kappa_{sc}}\mathcal{D}[\hat{a}_h]\hat{\rho}_{MC},
  \end{aligned}
\end{equation}
where in the last line we approximate $\mathcal{D}[\hat{a}_h]\hat{\rho}_{00}+\mathcal{D}[a_h^\dagger]\hat{\rho}_{11}]\approx\mathcal{D}[\hat{a}_h]\hat{\rho}$, as in the adiabatic regime the component $\hat{\rho}_{00}$ will be much more significant than $\hat{\rho}_{11}$. 

The form of Eq. \cref{eqSPGAhEliminated} matches that of the master equation Eq. \cref{eqMasterEquationSchrodingerApp}, and from it we can extract the parameters for the system with the dynamics of $\hat{b}$ eliminated. Abbreviating $\Omega=\frac{4g^2}{\kappa_{sc}}$:
\begin{equation}\label{eqAhEliminatedParameters}
  \begin{aligned}
    \hat{H}&=\Lambda(\hat{a}_e\hat{a}_h+\hat{a}_e^\dagger\hat{a}_h^\dagger), \\ 
    \hat{L} &= \sqrt{\Omega}\hat{a}_h.
  \end{aligned}
\end{equation}

We now perform a second adiabatic elimination, this time of $\hat{a}_h$, which will also be rapidly damped given its strong coupling to SC. Proceeding as before we derive
\begin{equation}
  \frac{d}{dt}\hat{\rho}_{20}\approx-\frac{\sqrt{2}\Lambda i}{\Omega}\hat{a}_e^\dagger\hat{\rho}_{10},
\end{equation}
where now $\hat{\rho}_{ij}$ act on $\mathcal{H}_{\hat{a}_e}$. Substituting this into the equation for $\frac{d}{dt}\hat{\rho}_{10}$ and considering the regime $\Omega\gg\Lambda$ leads to 
\begin{equation}
  \frac{d}{dt}\hat{\rho}_{10}=-\frac{2\Lambda i}{\Omega}(\hat{a}_e^\dagger\hat{\rho}_{00}-\hat{\rho}_{11}\hat{a}_e^\dagger),
\end{equation}
and hence, if we let $\hat{\rho}_e$ denote the density matrix of the $\hat{a}_e$ mode,
\begin{equation}\label{eqRhoDotHeralding}
  \frac{d}{dt}\hat{\rho}_e=\frac{\Lambda^2\kappa_{sc}}{g^2}\mathcal{D}[a_e^\dagger]\hat{\rho}_e.
\end{equation}

Comparing Eq. \cref{eqRhoDotHeralding} with Eq. \cref{eqMasterEquationSchrodinger}, we extract that in the regime $g,\kappa_{sc}\gg\Lambda$, $\kappa_{sc}\gg g$ and $\Omega=\frac{4g^2}{\kappa_{sc}}\gg\Lambda$, the evolution of $\hat{a}_e$ is described by the Hamiltonian and decoherence operator:
\begin{equation}\label{eqSPGFinalParams}
  \begin{aligned}
    \hat{H} &= 0, \\
    \hat{L} &= \sqrt{\gamma}\hat{a}_e^\dagger,
  \end{aligned}
\end{equation}
where we have defined $\gamma=\frac{\Lambda^2\kappa_{sc}}{g^2}$.

\bibliography{bibliography}

\end{document}